\documentclass{paradigmplus}
\usepackage[english]{babel}
\usepackage{makeidx}
\usepackage{graphicx}
\usepackage{hyperref}
\usepackage{fancyhdr}

\usepackage{verbatim} 
\usepackage{pifont}
\usepackage[mathscr]{euscript}
\usepackage{stmaryrd}
\usepackage{amssymb}
\usepackage{amsmath}
\usepackage{fancyvrb} 
\usepackage{mathrsfs}

\newtheorem{theorem}{Theorem}
\newtheorem{definition}[theorem]{Definition}

\title[A Workflow Language and Methodology (LSAWfP)]{A Language and Methodology based on Scenarios, Grammars and Views, for Administrative Business Processes Modelling}

\author[1,2,\correspondingAuthor]{Milliam Maxime {\textsc{Zekeng Ndadji}} \orcid{0000-0002-0417-5591}}
\author[1,2]{Maurice {\textsc{Tchoup\'e Tchendji}} \orcid{0000-0002-9208-6838}}
\author[1]{Cl\'ementin {\textsc{Tayou Djamegni}} \orcid{0000-0002-2231-3281}}
\author[3]{Didier \textsc{Parigot}}

\affil[1]{University of Dschang, Dschang, Cameroon \protect\\ \email{\{ndadji.maxime, maurice.tchoupe\}@univ-dschang.org, dtayou@yahoo.com}} 
\affil[2]{FUCHSIA Research Associated Team, \url{https://project.inria.fr/fuchsia/}}
\affil[3]{Inria, Sophia Antipolis, France \protect\\ \email{didier.parigot@inria.fr}} 

\keywords{Administrative Process Modelling, LSAWfP, Grammars, Artifact, Accreditation}

\graphicspath{{assets/}}

\vol{1}
\no{3}
\year{2020}
\doi{}
\publicationdates{2 September 2020}{24 October 2020}{25 October 2020}

\begin{document}
	\begin{abstract}
		In Business Process Management (BPM), process modelling has been solved in various ways. However, there are no commonly accepted modelling tools (languages). Some of them are criticized for their inability to capture both the lifecycle, informational and organizational models of processes. For some others, process modelling is generally done using a single graph; this does not facilitate modularity, maintenance and scalability. In addition, some of these languages are very general; hence, their application to specific domain processes (such as administrative processes) is very complex. 
		In this paper, we present a new language and a new methodology, dedicated to administrative process modelling. This language is based on a variant of attributed grammars and is able to capture the lifecycle, informational and organizational models of such processes. Also, it proposes a simple graphical formalism allowing to model each process's execution scenario as an annotated tree (modularity). In the new language, a particular emphasis is put on modelling (using "views") the perceptions that actors have on processes and their data. 
	\end{abstract}

\section{Introduction}
\label{Introduction}
Workflow technology also known as Business Process Management (BPM) technology, aims at automating business processes. A \textit{business process} is a set of tasks that follow a specific pattern and are executed to achieve a specific goal \cite{van2013business}. When such processes are managed electronically, they are called \textit{workflows}. 
To automate business processes, workflow technology provides a clear framework composed of two major entities: (1) a \textit{workflow language} for the description of such processes in a (generally graphical) format that can be interpreted by (2) a software system called \textit{Workflow Management System} (WfMS).
The role of WfMS is to facilitate collaboration and coordination of various actors involved in the distributed execution of processes' tasks: in this way, workflow technology reduces the automation of business processes to their modelling in \textit{workflow languages}; process modelling (specification) is therefore a crucial phase of workflow management\footnote{The \textit{Workflow Management Coalition} (it is the organization responsible for developing standards in workflow) defines \textit{workflow management} as the modelling and computer management of all the tasks and different actors involved in executing a business process \cite{van2013business}.}.

Several tools have been developed to address process modelling. Among the most well-known are the BPMN standard \cite{BPMN} and the YAWL language (\textit{Yet Another Workflow Language}) \cite{van2013business, van2005yawl}. 
Despite the significant research progress around these tools (often qualified as "\textit{traditional tools}"), they are not unanimously accepted. Indeed, they are often criticized for not being based on solid mathematical foundations \cite{borger2012approaches}, for having a much too great expressiveness compared to the needs of professionals in the field (this complicates their handling and increases the related costs) \cite{zur2013much} and/or for not being intuitive \cite{borger2012approaches}.

Another important criticism often levelled at traditional workflow languages is the fact that they treat data (process \textit{information model}) and users (part of process \textit{organizational model}) as second-class citizens by highlighting tasks and their routing (process \textit{lifecycle model}). To precisely remedy this, researchers have developed over the last two decades and under the initiative of IBM, the artifact-centric \cite{nigam2003business} approach to the design and execution of business processes. This one proposes a new approach to workflow management by focusing on both automated processes and data manipulated using the concept of "\textit{business artifact}" or "\textit{artifact}" in short.  
A major shortcoming of artifact-centric models is that, after designing a given business process, it's difficult to manage it out of the context for which it was designed: specification and execution contexts (the WfMS on which it must be executed) are strongly coupled. In fact, in artifact-centric approaches the process specification is done with artifact modelling and artifacts are usually tailored to dedicated collaborative systems; process designers are then obliged to take into account certain details related to the workflow execution technique during the modelling phase: it is therefore difficult to consider these approaches exclusively as business process modelling tools since they are execution-context dependant.

Another mentioned shortcoming of existing process modelling approaches is that they concentrate the modelling of a given process into a single task graph. This does not allow designers to explicitly express the entire control flow of certain types of processes; in addition, the resulting specifications are generally not easy to read, to maintain and to evolve. These concerns were first raised by Wil M. P. van der Aalst et al. \cite{van2001proclets, van2009workflow}. All these shortcomings of traditional workflow languages confirm that there is still a need of scientific innovation in the field of business process modelling. 

This paper presents a new \textit{Language for the Specification of Administrative Workflow Processes} (LSAWfP) based on the concept of attributed grammars. LSAWfP is built in a more traditional way and then, unlike the artifact-centric approaches, it allows process modelling independently of a workflow execution technique. Opposed to traditional workflow languages, LSAWfP provides coherent tools to model both processes' lifecycle model, information model and organizational model. LSAWfP is particularly interested in administrative process\footnote{These are processes for which the set of tasks (executed by humans or not) as well as their order of execution are known in advance \cite{mcCready, wilPetriNetWf}.} modelling as this type of process is the most frequently encountered in organizations \cite{mcCready, wilPetriNetWf}. 
A given administrative process is naturally composed of a set of execution scenarios; an \textit{execution scenario} (or simply \textit{scenario}) is an ordered subset of activities that, once executed, lead the process to one of its end states, whether or not a business goal is achieved.
In LSAWfP, the execution scenarios of a given administrative process are represented by a finite set $\left\{ \mathcal{S}_{ad}^1,\ldots,\mathcal{S}_{ad}^k \right\}$ of so-called \textit{representative scenarios} known in advance; representative scenario refers to any execution scenario that, in "combination" with some other representative scenarios, can generate a (potentially infinite) set of other scenarios (see sec. \ref{sec:TargetArtifacts}). Therefore, LSAWfP uses the \textit{scenario} as the modelling unit: a given process modelling consists to the modelling of each of its execution scenarios. Designers can thus focus on the modelling and the maintenance of process' parts rather than handling the whole process at a time: this seems to be more intuitive, modular and easier.

To use LSAWfP, the process to be modelled must be well understood by the designer; its tasks and their sequences, the data they produce and the actors taking part in their execution must be known in advance (administrative processes). In addition to these elements related to the lifecycle, the information model and the organizational model of the processes, the designer must be able to identify its various execution scenarios. The modelling approach (methodology) of LSAWfP can be described as follows: from the observation that one can analyse the textual description of a given administrative process to exhibit all its possible representative scenarios leading to its business goals, LSAWfP proposes to model each of these scenarios by an annotated tree called a \textit{representative artifact} in which, each node corresponds to a task of the process, and each hierarchical decomposition (a node and its sons) represents a scheduling of these tasks. From these representative artifacts, are derived an attributed grammar $\mathbb{G}$ called the \textit{Grammatical Model of Workflow} (GMWf). The symbols of a given GMWf represent the process tasks and each of its productions represents a scheduling of a subset of these tasks; intuitively, a production given by its left and right hand sides, specifies how the task on the left hand side precedes (must be executed before) those on the right hand side. Thus, the GMWf of a process contains both its \textit{information model} (modelled by its attributes) and its \textit{lifecycle model} (thanks to the set of its productions). Once the GMWf is obtained, LSAWfP propose to add organizational information (\textit{organizational model}) modelled by two lists: $\mathcal{L}_{P_k}$ which contains actors involved in the process and $\mathcal{L}_{\mathcal{A}_k}$ which contains their \textit{accreditations}. These lists aim at modelling actors, their roles and the different perceptions they have on a given process. Thus, with LSAWfP, the model (subsequently called \textit{a Grammatical Model of Administrative Workflow Process} - GMAWfP -) of a given administrative process $\mathcal{P}_{ad}$ is an executable grammatical specification given by a triplet 
$\mathbb{W}_f=\left(\mathbb{G}, \mathcal{L}_{P_k}, \mathcal{L}_{\mathcal{A}_k} \right)$.

The rest of this manuscript is organised as follows: after presenting some basic concepts, some related works and a running example (the peer-review process) in section \ref{sec:Préliminaire}, we present more formally and with illustrations, the proposed language in section \ref{sec:Contribution} and we discuss its expressiveness. 
A presentation of some ongoing works is conducted in section \ref{sec:Discussion}; in particular, we briefly present one of our current works that reinforces the justification of the need to produce a new workflow language. Finally, section \ref{sec:Conclusion} is devoted to the conclusion.

\section{Preliminaries and Related Works}
\label{sec:Préliminaire}
In this section, we present some basic concepts related to workflow technology to facilitate the understanding of this paper. We then give a very brief state of the art on process modelling techniques. We finally introduce a process that will be used for illustration purposes throughout this paper.

\subsection{Some Basic Concepts}
\label{sec:ConceptsDeBase}

\noindent\textbf{Workflow typology}: 
in the literature, there are several approaches to workflow classification.
However, it is the approach that classifies them by the nature and the behaviour of automated processes that is most commonly used. According to the latter, workflows are divided into three groups: production workflows, administrative workflows and ad-hoc workflows \cite{mcCready, wilPetriNetWf}. Production workflows are those automating highly structured processes that experience very little (or no) change over time. Administrative workflows apply to processes of which all cases are known; that means that tasks are predictable and their sequencing are simple and clearly defined. Ad-hoc workflows are more general; they automate occasional processes for which it is not always possible to define all the rules in advance.

The language presented in this paper is especially tailored for administrative workflows. One of the inherent characteristics of administrative business processes is the confidentiality that must sometimes be guaranteed on data and/or tasks that are executed. It is indeed easy to imagine administrative processes in which, various actors at any given time, have only a potentially partial perception of all the activities that have already and/or must be carried out: the perception that an actor has on the current state of a process is called his "view on the process". For example, in a peer-review process, a reviewer does not necessarily need to know if another reviewer has been contacted for the expertise of the article entrusted to him; and even if so, he should not necessarily know if the latter has already returned his report, etc. Similarly, when organising a journey for a Head of State, not all actors (secret services, civil office, doctor, presidential guard, etc.) have access to the same information which may include for example, tasks to be executed, their dates and states of execution, etc. Administrative workflows are characterized by the fact that all cases (tasks and their sequences), all actors and the permissions they have on tasks, etc. are known in advance. When specifying such processes, it should also be possible to model confidentiality constraints; for example, it should be possible to explicitly express the permissions which each actor has on each task. 
In LSAWfP, this requirement is treated as first-order concern with the help of a model called "accreditation", which allows to materialize the perception of each actor on the processes and their data.

~

\noindent\textbf{Business process specification}: 
the specification of a business process is commonly referred to as a \textit{workflow model}. According to \cite{divitini2001inter}, a workflow model consists of three main conceptual models: the \textit{organizational}, \textit{informational} and \textit{lifecycle} models. 
The \textit{organizational model} is used to express and classify the resources responsible for executing the tasks of the studied process. Generally, these are classified into \textit{roles} to which tasks are assigned. 
The \textit{informational model} is used to describe the structure of consumed and produced data during processes execution. 
Finally, the \textit{lifecycle model} is used to describe the structure of each task, the coordination between them and consequently, the coordination between the various actors involved in their execution. 
The lifecycle model is generally expressed using a language and allows the expression of basic control flows (\textit{sequential}, \textit{parallel}, \textit{alternative} and \textit{iterative}) between tasks. 
Ideally, a workflow language should be able to allow workflow model designers to express these three conceptual models.

\subsection{Related Works}
\label{sec:relatedWork}
There is a lot of work that has been already done in terms of process modelling. So, there is a plethora of workflow languages; the few presented here are intended to interest the reader. 

In the survey \cite{lu2007survey}, the authors group workflow languages into two categories: those based on graphical models (graph-based formalism), and the others based on rule specifications (rule-based formalism). In graph-based workflow languages, processes are specified using graphical models where tasks are represented as nodes, and control flow and data dependencies between tasks as arcs. In rule-based workflow languages, process logic is coded as a set of rules, each of which is associated with one or more business tasks and specifies their properties such as their pre and post conditions of execution.

BPMN and YAWL are graph-based workflow languages. The BPMN standard \cite{BPMN} was initiated by the \textit{Business Process Management Initiative} (BPMI) which merged with \textit{Object Management Group} (OMG) in 2005. It is a simple formalism inspired by the statecharts. BPMN is informal: i.e, it does not have well-defined semantic, so, resulting specifications are difficult to analyse \cite{borger2012approaches}. 
The YAWL language (\textit{Yet Another Workflow Language}) \cite{van2013business, van2005yawl} is based on a formalism called WF-Net (Workflow Net) \cite{wilPetriNetWf}, derived from that of Petri nets. Unlike BPMN, YAWL has a solid mathematical basis that facilitates the automatic analysis of its process models. 

As rule-based workflow languages, we can mention Event-Condition-Action (E-C-A) Business Rules \cite{knolmayer2000modeling} and ADEPT \cite{reichert2003adept}. The language in \cite{knolmayer2000modeling} is based on the E-C-A paradigm\footnote{E-C-A is a paradigm that specifies the desired behaviour for reactive systems (i.e. systems that maintain ongoing interactions with their environments). In such a system centered around the E-C-A paradigm, when an event occurs, a condition is evaluated (by a querying mechanism) and the system takes corresponding action \cite{almeida2005modular}.}; an E-C-A rule-based process model to serve as an integration layer between multiple process modelling languages is provided. In the ADEPT multi-agent system, process logic is expressed in the so-called service definition language (SDL); the resulting model is such as at runtime, agents have sufficient freedom to take alternative execution paths (from the model) to complete the process goal.

As mentioned in the introduction, for the last two decades, a lot of work on process modelling has been done on the artifact-centric paradigm \cite{nigam2003business, abi2016towards, deutsch2014automatic, hull2009facilitating, lohmann2010artifact, assaf2017continuous, assaf2018generating, boaz2013bizartifact, badouel14, badouel2015active, zekeng2020alanguage, zekeng2020lsawfp}. This paradigm was introduced by IBM through the work of Nigam and Caswell \cite{nigam2003business}. It recommends that, when modelling processes, one should focus on modelling a data structure called artifact that can give information both on the execution state of a process instance at a given time, and on the "how" to make this state evolve. 
Hull et al. \cite{hull2009facilitating} extends the artifact-centric model of \cite{nigam2003business} to provide an interoperation framework in which data are hosted on central infrastructures named \textit{artifact-centric hubs}. They propose mechanisms (including user views) for controlling access to these data. 
Lohmann and Wolf \cite{lohmann2010artifact} provide a choreography-like framework for artifact-centric interoperation. They abandon the fact of having a single artifact hub \cite{hull2009facilitating} and they introduce the idea of having several agents which operates on artifacts. Some of those artifacts are mobile; thus, the authors provide a systematic approach for modelling artifact location and its impact on the accessibility of actions using a Petri net. Badouel et al. \cite{badouel14, badouel2015active} introduce a flexible framework for data-centric case management. Their model puts stress on modelling process data and users as first class citizens. As for LSAWfP, they use an attributed grammar (named Guarded Attribute Grammar - GAG -) as the mathematical foundation of their model.

Some of the foundations of the artifact-centric paradigm come from the proclet model \cite{van2001proclets, van2009workflow}. In the latter, the authors provide a solution to the uniqueness of the task graph (which makes it unreadable and difficult to maintain) when modelling a given process. They introduce the concept of \textit{proclet}; they thus propose to deal with several levels of granularity assigned to lightweight workflow processes (proclets) in charge of orchestrating their execution. The modelling of each level of granularity is therefore done using a smaller task graph. We find this vision very interesting. However, the notion of granularity manipulated in \cite{van2001proclets} is not very intuitive and seems, as for artifact-centric models, intimately linked to the execution model of proclets. In the case of an administrative process $\mathcal{P}_{ad}$, we think it would be more affordable to partition its task graph according to a characteristic that is natural to it like its execution scenarios. Knowing that such a process is naturally composed of a set of execution scenarios and can be represented by a finite set $\left\{ \mathcal{S}_{ad}^1,\ldots,\mathcal{S}_{ad}^k \right\}$ of representative scenarios (see sec. \ref{sec:TargetArtifacts}) known in advance, we propose to use the scenario (execution scenario) as the modelling unit.

With the advent of cloud computing, several studies on business process execution have proposed completely decentralized models that can be deployed on Peer-to-Peer architectures \cite{fakas04, junYan06, huang2015cloud}. The most current solutions tend to have the processes executed by blockchain-based systems \cite{garcia2017optimized, carminati2018confidential, sturm2018lean, falazi2019modeling, lopez2019caterpillar, sturm2019blockchain, di2019blockchain}. There is therefore a need to adapt workflow languages so that the processes they specify can be executed in a distributed manner on such systems. The language we propose is in line with this need.

\subsection{A Running Example: the Peer-Review Process}
\label{sec:runningExample}

As running example, we will use the peer-review process. A brief description of it inspired by those made in \cite{van2001proclets, badouel14, zekeng2020alanguage, zekeng2020lsawfp}, can be the following one: 

\begin{itemize}
	\item The process starts when the editor in chief ($EC$) receives a paper for validation; 
	\item Then, the $EC$ performs a pre-validation after which he can accept or reject the submission for various reasons (subject of minor interest, submission not within the journal scope, non-compliant format, etc.); let us call this \textbf{task "$A$"};
	\item If he rejects the submission, he writes a report (\textbf{task "$B$"}) then notifies the corresponding author (\textbf{task "$D$"}) and the process ends;
	\item Otherwise, he chooses an associated editor ($AE$) and sends him the paper for the continuation of its validation;
	\item The $AE$ prepares the manuscript (\textbf{task "$C$"}) and contacts simultaneously two experts for the evaluation of the paper (\textbf{tasks "$E1$"} and \textbf{"$E2$"}); if a contacted expert refuses to participate, the $AE$ contacts another one (iteration on \textbf{task "$E1$"} or \textbf{"$E2$"}). Otherwise, the expert (referee) can start the evaluation;
	\item Each referee reads, seriously evaluates the paper (\textbf{tasks "$G1$"} and \textbf{"$G2$"}) and sends back a report (\textbf{tasks "$H1$"} and \textbf{"$H2$"}) and a message (\textbf{tasks "$I1$"} and \textbf{"$I2$"}) to the $AE$;
	\item After receiving reports from all referees, the $AE$ takes a decision and informs the $EC$ (\textbf{task "$F$"}) who sends the final decision to the corresponding author (\textbf{task "$D$"}).
\end{itemize} 

\begin{figure}[ht!]
	\noindent
	\makebox[\textwidth]{\includegraphics[scale=0.167]{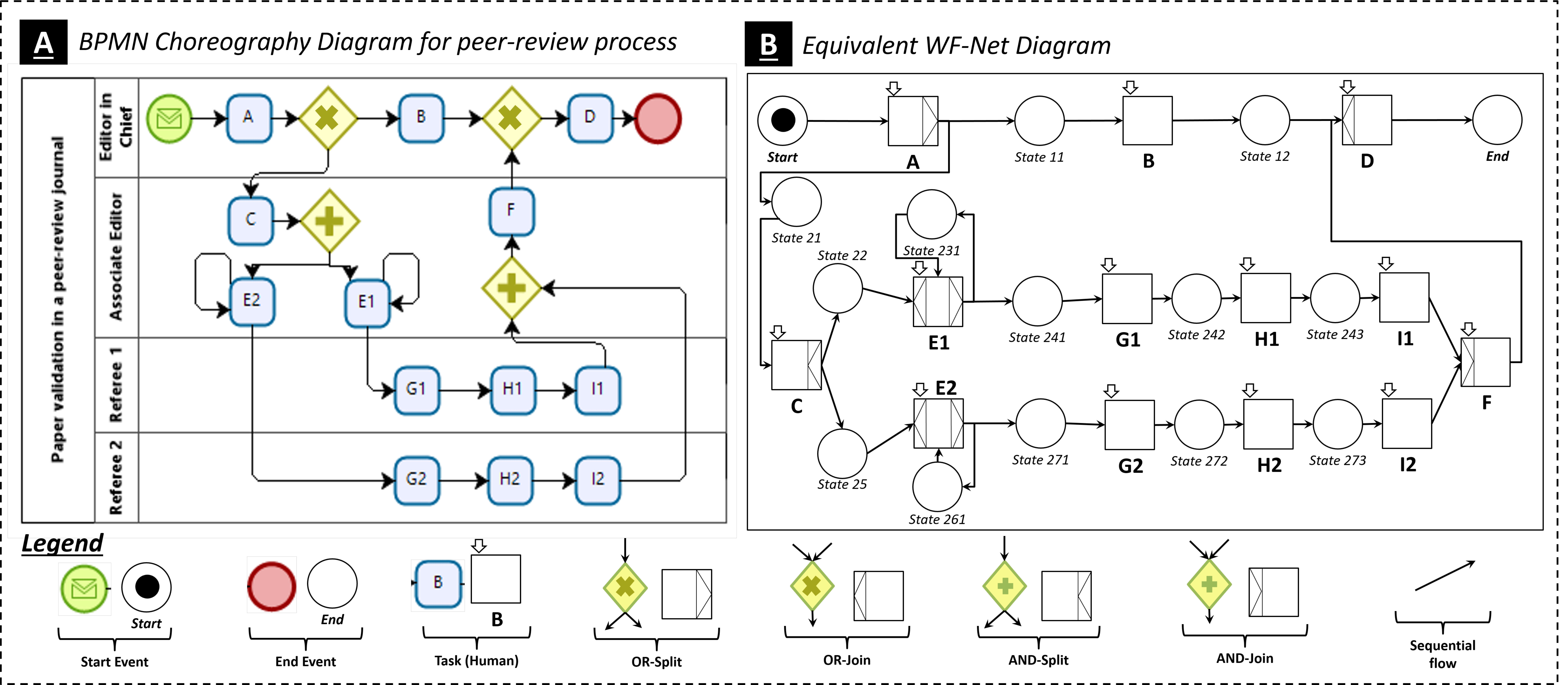}}
	\caption{Orchestration diagrams of the peer-review process.}
	\label{figPeerReviewProcess}
\end{figure}

From the description above, one can identify all the tasks to be executed, their sequencing, actors involved and the tasks assigned to them. For this case, four actors are involved: an editor in chief ($EC$) who is responsible for initiating the process, an associated editor ($AE$) and two referees ($R1$ and $R2$).
Figure \ref{figPeerReviewProcess} shows the orchestration diagrams corresponding to the graphical description of this peer-review process using BPMN (\textit{Business Process Model and Notation}) and WF-Net (\textit{Workflow Net}). Each diagram resumes the \textit{main scenarios} of the studied process. 
The purpose of this figure is to show the independence of our running example from a specific workflow language; in addition, we want the reader to bear in mind what the two workflow languages used in this figure offer in terms of process modelling tools and methodology in order to better understand the contributions of LSAWfP. It can be seen that with both formalisms, the modelling of all scenarios is tackled at the same time and process data is treated in background or not at all. With BPMN, the organizational aspect highlights the actors and the tasks assigned to them, but nothing more. The organizational aspect is absent with the WF-Net formalism; however, the control flow is more expressive because it highlights the different states of the process after the execution of each task. We have added a small caption for the uninformed reader; the bibliographical references on the used languages can also be a valuable aid to the understanding of this figure.

\section{A Language for the Specification of Administrative Workflow Processes (LSAWfP)}
\label{sec:Contribution}
In this section, we present the new language LSAWfP that allows to specify administrative workflow processes independently of a workflow execution technique, and with the use of scenario as modelling unit.

\subsection{Artifacts as Control Flow Graphs}
\label{sec:ArtifactsStructure}
Let's consider an administrative process $\mathcal{P}_{ad}$ to be modelled. By definition (of administrative process), its set $\mathbb{T}_n = \{X_1, \ldots, X_n\}$ of tasks is known in advance. In traditional workflow languages like BPMN or WF-Net, the control flow between its tasks is represented using a directed graph that can contain cycles (see fig. \ref{figPeerReviewProcess}). Such a graph allows the modelling of the potentially infinite set\footnote{This is the case when there is one or more iterative routing (materialized by cycles in the task graph) on tasks.} of $\mathcal{P}_{ad}$'s execution scenarios.
Let's note however that  each $\mathcal{P}_{ad}$'s execution scenario can also be modelled using an annotated tree $t_i$ called \textit{artifact}. Indeed, starting from the fact that a given scenario $\mathcal{S}_{ad}^i$ consists of a subset $\mathbb{T}_m \subseteq \mathbb{T}_n$ of $m \leq n$ tasks whose instances are to be executed in a specific order (in parallel or in sequence), one can represent $\mathcal{S}_{ad}^i$ as a tree $t_i$ in which each node (a task instance labelled $X_i$) potentially corresponds to a task $X_i \in \mathbb{T}_m$ of $\mathcal{S}_{ad}^i$ and each hierarchical decomposition (a node and its sons) corresponds to a scheduling: the task associated with the parent node must be executed before those associated with the son nodes; the latter must be executed according to an order - parallel or sequential - that can be specified by particular annotations "$\fatsemi$" (is sequential to) and "$\parallel$" (is parallel to) which will be applied to each hierarchical decomposition. The annotation "$\fatsemi$" (resp. "$\parallel$") reflects the fact that the tasks associated with the son nodes of the decomposition must (resp. can) be executed in sequence (resp. in parallel). To model iteration, nodes can be recursive in an artifact: i.e a node labelled $X_i$ may appear in subtrees rooted by a node having the same label $X_i$.

\begin{figure}[ht!]
	\noindent
	\makebox[\textwidth]{\includegraphics[scale=0.25]{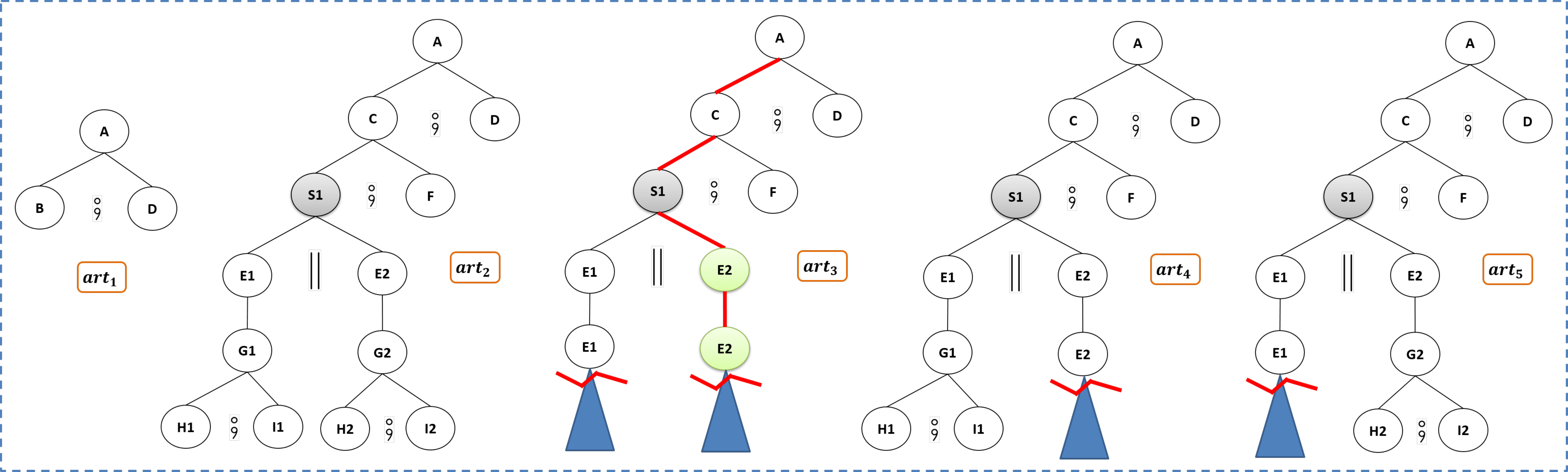}}
	\caption{Representative artifacts of a paper validation process in a peer-review journal.}
	\label{fig:artefactsGlobaux}
\end{figure}

Considering the running example (the peer-review process), two of its execution scenarios can be modelled using the two artifacts $art_1$ and $art_2$ in figure \ref{fig:artefactsGlobaux}. In particular, we can see that $art_1$ shows how the task "Receipt and pre-validation of a submitted paper" assigned to the $EC$, and associated with the symbol $A$ (see sec. \ref{sec:runningExample}), must be executed before tasks associated with the symbols $B$ and $D$ that are to be executed in sequence from the left to the right. Note that additional symbols called \textit{(re)-structuring symbols} can be added in artifacts to correct the scheduling of tasks (we better explain this in section \ref{sec:GMWf}): this is the case for $art_2$ in which the symbol $S1$ has been added.

\subsection{Representative Artifacts and Grammatical Model of Workflow}
\label{sec:TargetArtifactsAndGMWf}

\subsubsection{Representative Artifacts}
\label{sec:TargetArtifacts}
As mentioned earlier (see sec. \ref{sec:ArtifactsStructure}), the set of execution scenarios for a given administrative process can be infinite. This is the case of our running example process in which we can iterate on tasks $E1$ and $E2$ without limit; as each iteration on either $E1$ or $E2$ give rise to a new execution scenario, we thus generate an infinite set of execution scenarios. In these cases, the designer cannot list this set of scenarios in order to model each of them. This problem can be avoided by taking inspiration from the role played by the concept of vector spaces' basis in mathematics (algebra). In this sense, we can find a finite set $\tau=\left\{ \mathcal{S}_{ad}^1,\ldots,\mathcal{S}_{ad}^k \right\}$ of scenarios said to be \textit{representative} and modelled be a finite set $\zeta=\left\{t_1,\ldots,t_k\right\}$ of \textit{representative artifacts} such that, any artifact representing an execution scenario can be expressed as a "combination" of some elements of $\zeta$. 

We designate by the expression \textit{nominal scenario} of a given process, any scenario leading to a given business goal without iteration; in the same vein, scenarios in which at least one iteration have been made are called \textit{alternative scenarios}. For a given process, the set of nominal scenarios is finite and the artifacts depicting each of these scenarios are part of the process representative artifacts. The other part of the process representative artifacts is determined much more technically, from the (potentially infinite) set of alternative scenarios. Concretely, when designing an alternative scenario artifact, the designer must prune it at each first iteration encountered: i.e, the designer must prune each branch of an alternative scenario artifact as soon as he encounters a node labelled for the second time, by a same label along a path starting from the root. By doing so, the designer does not redevelop a node into a subtree that has already been explored: thus, he does not loop endlessly on potential iterations and does not explore all alternative scenarios generated by those iterations. However, the resulting pruned artifact contains patterns (productions) that indicate how to construct the artifacts associated with the considered alternative scenarios: this is why the obtained pruned artifact is said to be representative. 

More precisely, one could assume that to design the representative artifacts of a given business process, the designer begins by identifying the initial tasks of it (i.e., the tasks that can start one of its execution scenarios); each of these tasks will thus constitute the root of several representative artifacts. To construct the set $arts_{X_{0_{i}}}$ of representative artifacts rooted in a given initial task $X_{0_{i}}$, the designer will:
\begin{enumerate}
	\item[(1)] Construct an artifact $art$ having $X_{0_{i}}$ as the single node (root);
	\item[(2)] Then, he will determine the set $follow=\left\{\left(X_{1_{i_{1}}}, \ldots, X_{m1_{i_{1}}}\right), \ldots, \left(X_{1_{i_{n}}}, \ldots, X_{mn_{i_{n}}}\right)\right\}$ of task combinations (each combination is either sequential or parallel\footnote{If a given combination $\left(X_{1_{i_{j}}}, \ldots, X_{mj_{i_{j}}}\right)$ is sequential (resp. parallel), its tasks are to be (resp. can be) executed sequentially (resp. in parallel).}) that can be immediately executed after the execution of $X_{0_{i}}$. For each combination $\left(X_{1_{i_{j}}}, \ldots, X_{mj_{i_{j}}}\right)$, the designer will replace the artifact $art$ by a new artifact $art_j$ obtained by expanding the node $X_{0_{i}}$ of $art$ such that in $art_j$, the tasks $X_{1_{i_{j}}}, \ldots, X_{mj_{i_{j}}}$ are the child nodes of $X_{0_{i}}$.
	\item[(3)] It will then only remain to recursively develop (using the principle of (2)) each leaf node of the new artifacts until representative artifacts (those that describe an execution scenario in its entirety) are obtained.
\end{enumerate}

This construction principle emphasizes the fact that one does not loose information by pruning an artifact when encountering a given node $X$ for the second time in the same branch. In such a case, it is not necessary to develop $X$ a second time since the designer has enumerated (in several artifacts) all the possibilities (scenarios) of continuing the execution of the process after the execution of the task associated with $X$. As we will see in section \ref{sec:GMWf}, these possibilities will be coded in a grammar and thus, the execution scenarios characterized by several iterations on $X$, will indeed be specified in the language. When constructing a representative artifact, the pruning of a branch is therefore systematic when a node is encountered for the second time; no matter how many nodes generate an iteration in the same branch.

Figure \ref{fig:artefactsGlobaux} presents the five representative artifacts of our running example process. The artifacts $art_1$ and $art_2$ model the two nominal scenarios: $art_1$ models the scenario in which the $EC$ directly rejects the paper while $art_2$ models the case where the paper is evaluated by referees ($R1$ and $R2$) without the $AE$ having to contact more than two experts (no iteration on tasks $E1$ and $E2$). The artifacts $art_3$, $art_4$ and $art_5$ represent the infinite set of alternative scenarios in this example: some of their subtrees (those represented by blue triangles) have been pruned. For illustration purposes, we put forward the pruning made by the designer on the node $E2$ (in green colour) of $art_3$, which appeared for the second time in the same branch (the branch is highlighted in red colour).

\subsubsection{Grammatical Model of Workflow}
\label{sec:GMWf}
From the finite set of representative artifacts of a given process, it is possible to extract an abstract grammar\footnote{It is enough to consider the set of representative artifacts as a "set of generators" of a regular tree language: there is therefore an (abstract) grammar to generate them.} that represents the underlying process's lifecycle model : it is this grammar that we designate by the expression \textit{Grammatical Model of Workflow (GMWf)}.

Let's consider the set $\left\{t_1,\ldots,t_k\right\}$ of representative artifacts modelling the $k$ representative scenarios of a given process $\mathcal{P}_{ad}$ of $n$ tasks ($\mathbb{T}_n = \{X_1, \ldots, X_n\}$). Each $t_i$ is a derivation tree for an abstract grammar (a GMWf) $\mathbb{G}=\left(\mathcal{S},\mathcal{P},\mathcal{A}\right)$ whose set of symbols is $\mathcal{S}=\mathbb{T}_n$ (all process tasks) and each production $p \in \mathcal{P}$ reflects a hierarchical decomposition contained in at least one of the representative artifacts. Each production is therefore exclusively of one of the following two forms: $p: X_0 \rightarrow X_1 \fatsemi \ldots \fatsemi X_n$ or  $p: X_0 \rightarrow X_1 \parallel \ldots \parallel X_n$. The first form $p: X_0 \rightarrow X_1 \fatsemi \ldots \fatsemi X_n$ (resp. the second form $p: X_0 \rightarrow X_1 \parallel \ldots \parallel X_n$) means that task $X_0$ must be executed before tasks $\left\{X_1,\ldots,X_n\right\}$ that must be (resp. can be) executed in sequence (resp. in parallel) from the left to the right. A GMWf can therefore be formally defined as follows:
\begin{definition}
	\label{defGMWf1}
	A \textbf{Grammatical Model of Workflow} (GMWf) is defined by $\mathbb{G}=\left(\mathcal{S},\mathcal{P},\mathcal{A}\right)$
	where :
	\begin{itemize}
		\item $\mathcal{S}$ is a finite set of \textbf{grammatical symbols} or \textbf{sorts} corresponding to various \textbf{tasks} to be executed in the studied business process; 
		\item $\mathcal{A}\subseteq \mathcal{S}$ is a finite set of particular symbols called \textbf{axioms}, representing tasks that can start an execution scenario (roots of representative artifacts), and 
		\item $\mathcal{P}\subseteq\mathcal{S}\times\mathcal{S}^{*}$ is a finite set of \textbf{productions} decorated by the annotations "$\fatsemi$" (is sequential to) and "$\parallel$" (is parallel to): they are \textbf{precedence rules}. 
		A production $P=\left(X_{P(0)},X_{P(1)},\cdots, X_{P(|P|)}\right)$ is either of the form $P: X_0 \rightarrow X_1 \fatsemi \ldots \fatsemi X_{|P|}$, or of the form $P: X_0 \rightarrow X_1 \parallel \ldots \parallel X_{|P|}$ and $\left|P\right|$ 
		designates the length of $P$'s right-hand side.
		A production with the symbol $X$ as left-hand side is called a \textit{X-production}.
	\end{itemize}
\end{definition}

Let's illustrate the notion of GMWf by considering the one generated from an interpretation of the representative artifacts for the peer-review process (see fig. \ref{fig:artefactsGlobaux}): the derived GMWf is  $\mathbb{G}=\left(\mathcal{S},\mathcal{P},\mathcal{A}\right)$ in which the set $\mathcal{S}$ of grammatical symbols is
$\mathcal{S}=\{A, B, C, D, S1, E1, E2, F, G1, G2, H1, H2, I1, I2\}$ (see sec \ref{sec:runningExample});
the only initial task (axiom) is $A$ (then $\mathcal{A}=\{A\}$) and the set $\mathcal{P}$ of productions is\footnote{A production of the form  $X\rightarrow \varepsilon$ indicates that task $X$ is not "decomposable" in subtasks.}:
\[ 
\begin{array}{l|l|l|l}
P_{1}:\; A\rightarrow B\fatsemi D & \; P_{2}:\; A\rightarrow C\fatsemi D\; & \; P_{3}:\; C\rightarrow S1\fatsemi F\; & \; P_{4}:\; S1\rightarrow E1\parallel E2 \\
P_{5}:\; E1\rightarrow G1 & \; P_{6}:\; E2\rightarrow G2\; & \; P_{7}:\; E1\rightarrow E1\; & \; P_{8}:\; E2\rightarrow E2 \\
P_{9}:\; G1\rightarrow H1 \fatsemi I1 & \; P_{10}:\; G2\rightarrow H2 \fatsemi I2\; & \; P_{11}:\; B\rightarrow \varepsilon\; & \; P_{12}:\; D\rightarrow \varepsilon  \\
P_{13}:\; F\rightarrow \varepsilon & \; P_{14}:\; H1\rightarrow \varepsilon\; & \; P_{15}:\; I1\rightarrow \varepsilon \; & \; P_{16}:\; H2\rightarrow \varepsilon \\ 
P_{17}:\; I2\rightarrow \varepsilon & & & \\
\end{array}
\]

There may be special cases where it is not possible to schedule the tasks of a scenario using the two (only) forms of production selected for GMWf. For example, this is the case for the peer-review process wherein task $C$ precedes tasks $E1$, $E2$ and $F$, tasks $E1$ and $E2$ can be executed in parallel and precede $F$ (see sec. \ref{sec:runningExample}). 
In such cases, the introduction of a few new symbols known as \textit{(re)structuring symbols} (not associated with tasks) can make it possible to produce a correct scheduling. For the peer-review process example, the introduction of a new symbol $S1$ allows us to obtain the following productions: $P_{3}: C\rightarrow S1\fatsemi F$ and $P_{4}:\; S1\rightarrow E1\parallel E2$ which properly model the required scheduling and avoid the usage of the malformed production $p:C \rightarrow E1 \parallel E2 \fatsemi F$ (see in fig. \ref{fig:artefactsGlobaux}, $art_2$, the node $S1$ --- in gray ---). 
To deal with such cases, the previously given GMWf definition (definition \ref{defGMWf1}) is slightly adapted by integrating the (re)structuring symbols; the resulting definition is as follows:
\begin{definition} 
	\label{defGMWf2}
	A \textbf{Grammatical Model of Workflow} (GMWf) is defined by $\mathbb{G}=\left(\mathcal{S},\mathcal{P},\mathcal{A}\right)$
	wherein 
	$\mathcal{P}$ and $\mathcal{A}$ refer to the same purpose as in definition \ref{defGMWf1}, 
	$\mathcal{S}=\mathcal{T} \cup \mathcal{T}_{Struc}$ 
	is a finite set of \textbf{grammatical symbols} or \textbf{sorts} in which, those of $\mathcal{T}$ correspond to \textbf{tasks} of the studied business process, while those of $\mathcal{T}_{Struc}$ are (re)structuring symbols.
\end{definition}

\subsection{Modelling the Information and Organization Model of Processes with LSAWfP}

\subsubsection{An Information Model for LSAWfP}
\label{sec:GMWfInformationModel}
As formalized in definition \ref{defGMWf2}, a GMWf perfectly models the tasks and control flow of administrative processes (lifecycle model). In this section we discuss the specification of processes-related data (\textit{the information model}) in LSAWfP.

It is not easy to model the structure of business processes data using a general type as they differ from one process to another. For the current work, tackling the processes data structure has no proven interest because it does not bring any added value to the proposed model since, we are not specifically interested in data modelling but rather in process modelling: a representation of these data using a set of variables associated with tasks is largely sufficient. However, it should be noted that in existing data-driven modelling approaches like the Guarded Attribute Grammar (GAG) model \cite{badouel2015active, theseNsaibirni2019, tchoupetchendji:hal-02925745}, each task comes equipped with a set of \textit{inherited} attributes (terms over a ranked alphabet) and a set of \textit{synthesised} attributes where: inherited attributes represents input data (i.e, necessary data for the associated task to be executed) while synthesised attributes represents output data (i.e, data that are produced after the task being executed). In addition, dependency relationships between data (attributes) are often specified.

In this work, the potentially manipulated data by a given process task is represented using a single \textit{attribute} embedded in the nodes associated with it. This is more than enough to show that LSAWfP cares about processes' data (this is one of this paper's goals). In more specific (future) work, the form of this attribute can be simply refined (as in the GAG approach \cite{badouel2015active, theseNsaibirni2019, tchoupetchendji:hal-02925745}) to allow designers to better describe the nature of the manipulated data and their impact on processes. To formalize the taking into account of attributes, we update for the last time the definition of GMWf. We thus associate with each symbol, an attribute named $status$ allowing to store all the data of the associated task; its precise type is left to the discretion of the process designer. The new definition of GMWf is thus the following one:
\begin{definition} 
	\label{defGMWf3}
	A \textbf{Grammatical Model of Workflow} (GMWf) is defined by $\mathbb{G}=\left(\mathcal{S},\mathcal{P},\mathcal{A}\right)$
	wherein 
	$\mathcal{S}$, $\mathcal{P}$ and $\mathcal{A}$ refer to the same purpose as in definition \ref{defGMWf2}.
	Each grammatical symbol $X\in\mathcal{S}$ is associated with an attribute named \textbf{\textit{status}}, that can be updated when tasks are executed; $\textbf{X.status}$ provides access (read and write) to its content.
\end{definition}


\subsubsection{An Organizational Model for LSAWfP}
\label{sec:GMWfOrganizationalModel}
Because business processes are generally carried out collectively, it is important to model actors and to set up mechanisms to ensure better coordination between them and to eventually guarantee the confidentiality of certain actions and data: this is the purpose of \textit{accreditation}.
The accreditation of a given actor provides information on its rights (permissions) relatively to each sort (task) of the studied process's GMWf. 
We propose here, a simple but non-exhaustive nomenclature of rights. It is inspired by the one used in UNIX-like operating systems. Three types of accreditation are therefore defined: accreditation in reading \textit{(r)}, writing \textit{(w)} and execution \textit{(x)}. 

\begin{enumerate}
\item[\textbf{1.}] \textit{The accreditation in reading \textit{(r)}}: an actor accredited in reading on sort $X$ must be informed of the execution of the associated task; he must also have free access to its execution state (data generated during its execution).
We call an actor's \textbf{\textit{view}}, the set of sorts on which he is accredited in reading.
\item[\textbf{2.}] \textit{The accreditation in writing \textit{(w)}}: an actor accredited in writing on sort $X$ can execute the associated task. The designation of the right to execute a task by the term accreditation in writing can be confusing. However, we consider that the execution of tasks is performed (manually and/or automatically) by human actors. At the end of a given task execution, the actor in charge of its execution must enter (write) the produced data into the system: he must have an accreditation in writing. To make it simple, any actor accredited in writing on a sort must necessarily be accredited in reading on it\footnote{This hypothesis, which does not reduce the expressiveness of the language, is taken only because it is estimated that actors will operate through What You See Is What You Get (WYSIWYG) tools.}. 
\item[\textbf{3.}] \textit{The accreditation in execution \textit{(x)}}: an actor accredited in execution on sort $X$ is allowed to ask the actor who is accredited in writing in it, to execute it (realization of the associated task). This right is particularly appropriate for the modelling of interaction between actors: especially in the case of processes where it is important to know "who" can ask "who" to perform a given task. This is not a "delegation" of work, since there is no transfer of tasks, or of competencies. Work delegation can be the subject of more elaborate work on the LSAWfP organizational model.
\end{enumerate}
More formally, an accreditation is defined as follows:
\begin{definition} 
	\label{defSyllabaire}
	An \textbf{accreditation} $\mathcal{A}_{A_i}$ defined on the set $\mathcal{S}$ of grammatical symbols for an actor $A_i$, is a triplet $\mathcal{A}_{A_i}=\left(\mathcal{A}_{A_i(r)},\mathcal{A}_{A_i(w)},\mathcal{A}_{A_i(x)}\right)$ such that, 
	$\mathcal{A}_{A_i(r)} \subseteq \mathcal{S}$ also called \textbf{view} of actor $A_i$, is the set of symbols on which $A_i$ is accredited in reading, 
	$\mathcal{A}_{A_i(w)} \subseteq \mathcal{A}_{A_i(r)}$ is the set of symbols on which $A_i$ is accredited in writing and  
	$\mathcal{A}_{A_i(x)} \subseteq \mathcal{S}$ is the set of symbols on which $A_i$ is accredited in execution.
\end{definition}

The accreditations of various actors must be produced by the workflow designer just after modelling the scenarios in the form of representative artifacts. From the task assignment for the peer-review process in the running example (see sec. \ref{sec:runningExample}), it follows that the accreditation in writing of the $EC$ is $\mathcal{A}_{EC(w)}=\{A, B, D\}$, that of the $AE$ is $\mathcal{A}_{AE(w)}=\{C, S1, E1, E2, F\}$ and that of the first (resp. the second) referee is $\mathcal{A}_{R_1(w)}=\{G1, H1, I1\}$ (resp. $\mathcal{A}_{R_2(w)}=\{G2, H2, I2\}$).
Since the $EC$ can only execute the task $D$ if the task $C$ is already executed (see fig. \ref{fig:artefactsGlobaux}), in order for the $EC$ to be able to ask the $AE$ to execute this task, he must be accredited in execution on it; so we have $\mathcal{A}_{EC(x)}=\{C\}$.
Moreover, in order to be able to access all the information on the peer-review evaluation of a paper (task $C$) and to summarize the right decision to send to the author, the $EC$ must be able to consult the reports (tasks $I1$ and $I2$) and the messages (tasks $H1$ and $H2$) of the different referees, as well as the final decision taken by the $AE$ (task $F$). These tasks, added to $\mathcal{A}_{EC(w)}$\footnote{Recall that we consider that one can only execute what he sees.} constitute the set $\mathcal{A}_{EC(r)}=\mathcal{V}_{EC}=\{A, B, C, D, H1, H2, I1, I2, F\}$ of tasks on which he is accredited in reading. By doing so for each of other actors, we deduce the accreditations represented in table \ref{tableau:vuesActeurs}.
\begin{table}[ht]
	\centering
	\caption{Accreditations of the different actors taking part in the peer-review process.}
	\label{tableau:vuesActeurs}
	\begin{tabular}[t]{p{1.2cm}p{11.6cm}}
		\hline \hline
		\textbf{Actor} & \textbf{Accreditation} \\
		\hline
		$EC$ & $\mathcal{A}_{EC}=\left(\{A, B, C, D, H1, H2, I1, I2, F\}, \{A, B, D\}, \{C\}\right)$ \\
		\hline
		$AE$ & $\mathcal{A}_{AE}=\left(\{A, C, S1, E1, E2, F, H1, H2, I1, I2\}, \{C, S1, E1, E2, F\}, \{G1, G2\}\right)$ \\
		\hline
		$R1$ & $\mathcal{A}_{R1}=\left(\{C, G1, H1, I1\}, \{G1, H1, I1\}, \emptyset\right)$ \\
		\hline
		$R2$ & $\mathcal{A}_{R2}=\left(\{C, G2, H2, I2\}, \{G2, H2, I2\}, \emptyset\right)$ \\
		\hline \hline
	\end{tabular}
\end{table}

Since the (re)structuring symbols are not associated with tasks and were only introduced to adjust the control flow, their execution neither requires nor produces data; they play the same role as gateways in traditional workflow languages. Therefore, the accreditation in writing and execution on them may be best left to the designer's appreciation; he will then make the assignment by referring to the execution model he will use later. To this end, he could use the same principle for the assignment of these accreditations in the case of concrete process' tasks. However, one could by default consider that all actors are accredited in reading on (re)structuring symbols; this would make these symbols visible to all of them and would guarantee that the adjustment of the control flow will be effective for all of them even if they have partial perceptions of the process.

\subsection{Summary}
\label{sec:LSAWfP}

\subsubsection{Definition of LSAWfP}
To summarise, we state that in LSAWfP, an administrative process $\mathcal{P}_{ad}$ is specified using a triplet $\mathbb{W}_f=\left(\mathbb{G}, \mathcal{L}_{P_k}, \mathcal{L}_{\mathcal{A}_k} \right)$ called \textit{a Grammatical Model of Administrative Workflow Process} (GMAWfP) and composed of: 
a GMWf, a list of actors and a list of their accreditations. 
The GMWf is used to describe all the tasks of the studied process and their scheduling, while the list of accreditations provides information on the role played by each actor involved in the process execution.  
A GMAWfP can then be formally defined as follows:
\begin{definition}
	\label{defMGSPWfA}
	A \textbf{Grammatical Model of Administrative Workflow Process} (GMAWfP) $\mathbb{W}_f$ for a given business process, is a triplet $\mathbb{W}_f=\left(\mathbb{G}, \mathcal{L}_{P_k}, \mathcal{L}_{\mathcal{A}_k} \right)$
	wherein $\mathbb{G}$ is the studied process (global) GMWf, $\mathcal{L}_{P_k}$ is the set of $k$ actors taking part in its execution and $\mathcal{L}_{\mathcal{A}_k}$ represents the set of these actors accreditations. 
\end{definition}

\subsubsection{How to Model a Process with LSAWfP (Methodology)}
\begin{figure}[ht!]
	\noindent
	\makebox[\textwidth]{\includegraphics[scale=0.27]{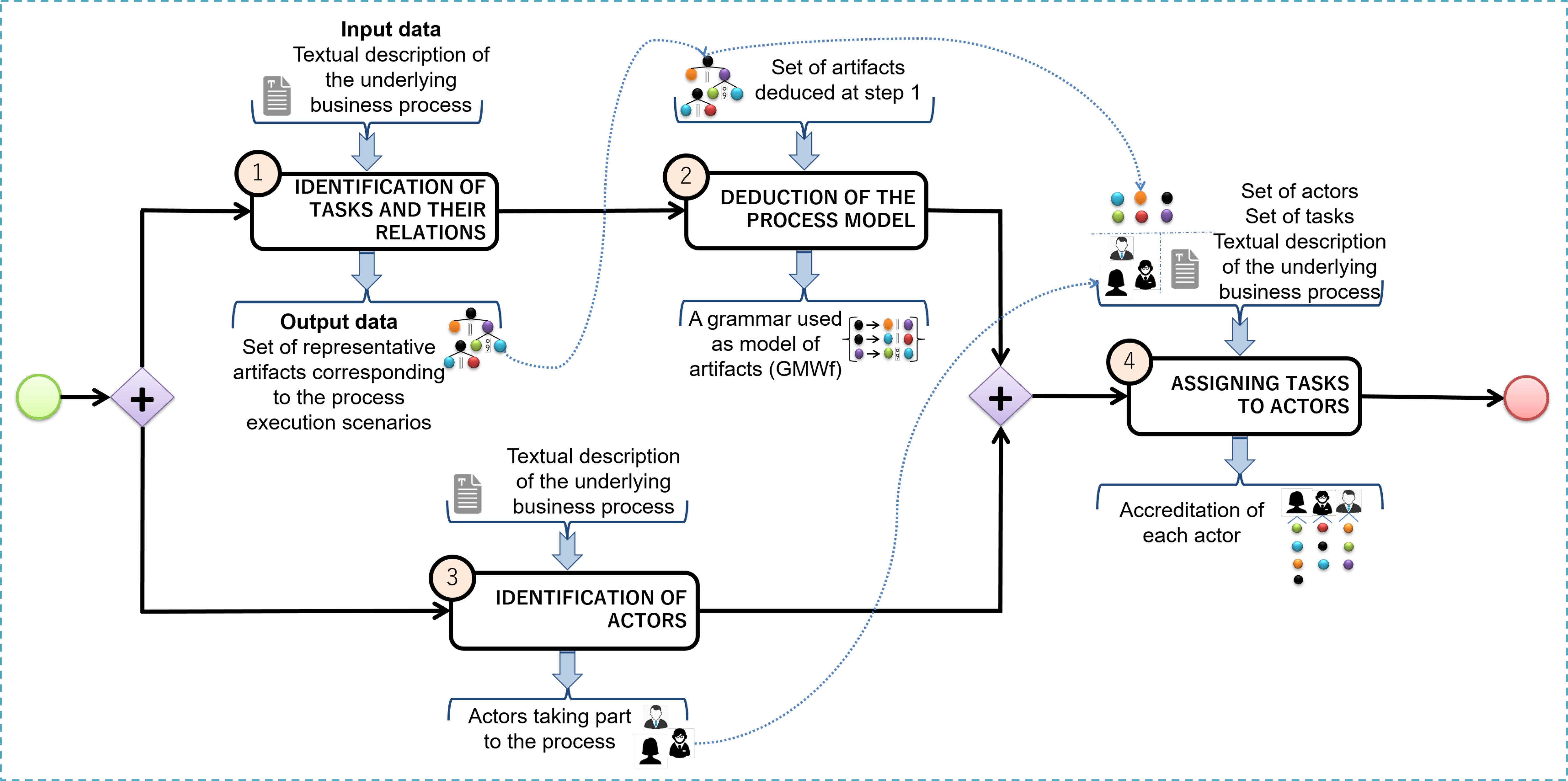}}
	\caption{The main activities that are carried out when modelling a process using LSAWfP.}
	\label{fig:methodology}
\end{figure}

To model a given process using LSAWfP, one must start from a textual description of the process and perform the four activities illustrated in figure \ref{fig:methodology}. First, the set of tasks and their execution precedence relationships must be identified in order to produce a finite set of representative artifacts following the approach presented in section \ref{sec:TargetArtifacts}. The GMWf must then be deduced from the set of representative artifacts thus produced. Then, the different actors involved in the execution of the process being modelled must be identified, and finally, a coherent assignment of tasks to actors must be made and their respective accreditations deduced.

\subsection{On the Expressiveness of LSAWfP}
\label{sec:Expressiveness}
Let's consider a specification  $\mathbb{W}_f=\left(\mathbb{G}, \mathcal{L}_{P_k}, \mathcal{L}_{\mathcal{A}_k} \right)$ of a given business process $\mathcal{P}_{ad}$. 
As described above, its organizational model that expresses and classifies/assigns the resources that must execute its tasks is given by the couple $\left( \mathcal{L}_{P_k}, \mathcal{L}_{\mathcal{A}_k} \right)$ of $\mathbb{W}_f$.
Its informational model that describes the data structure being manipulated is given by the type of the attribute \textit{status} associated with each task. 
Its lifecycle model that provides information on tasks and their sequencing (coordination) is given by the GMWf $\mathbb{G}$ of $\mathbb{W}_f$. Thus, we can conclude that LSAWfP has the major expected characteristics of a workflow language according to \cite{van2013business}. 

\begin{figure}[ht!]
	\begin{center}
		\includegraphics[scale=0.25]{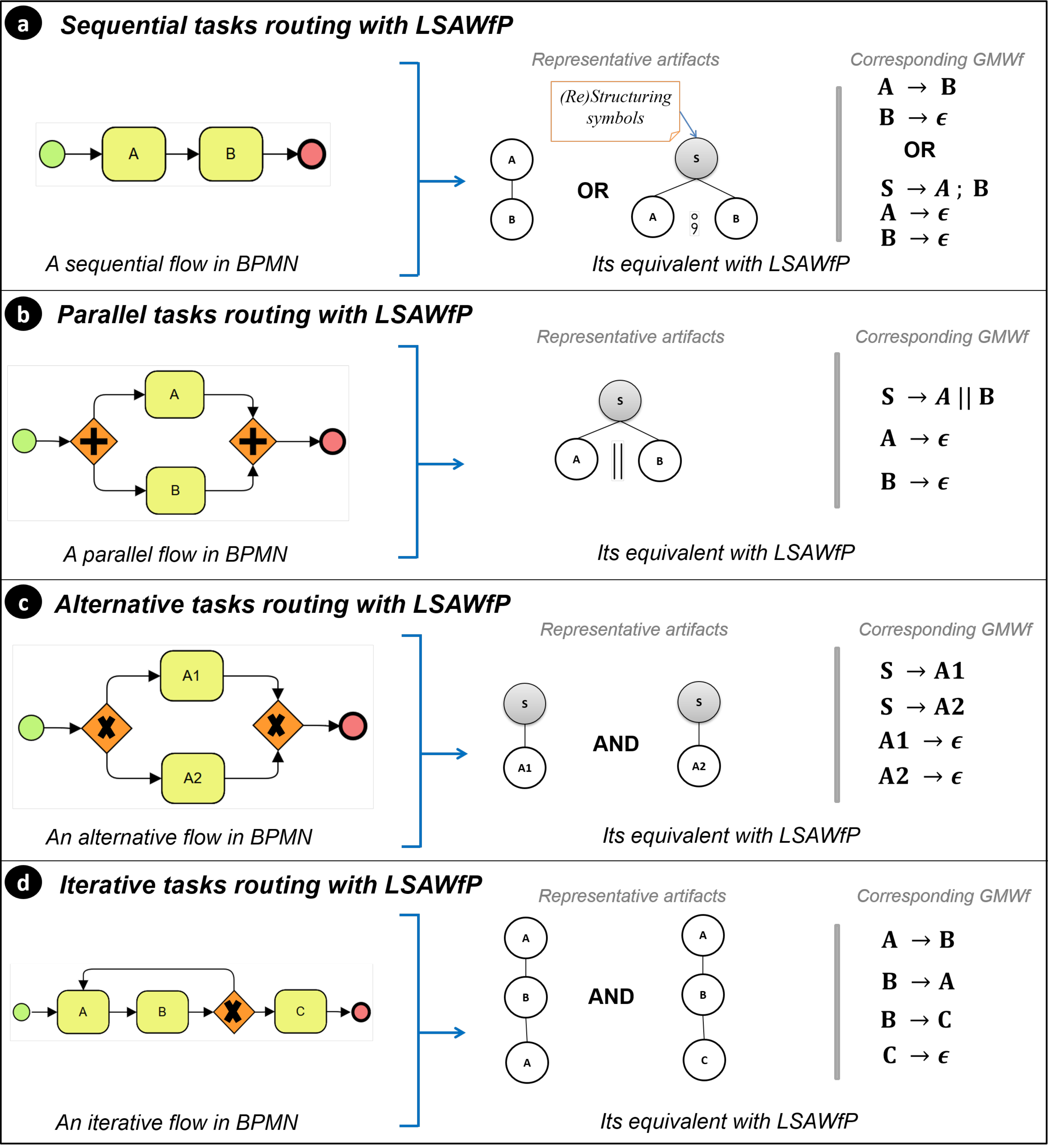}
		\caption{Illustrating basic control flows with LSAWfP.}
		\label{routingWithLSPWfA}
	\end{center}
\end{figure}

The GMWf effectively allows the designers to specify all the basic control flows (sequential, parallel, alternative and iterative) which can be found in traditional workflow languages.
Figure \ref{routingWithLSPWfA} gives for each type of basic control flow its BPMN notation and the corresponding notations (artifact and associated productions) in LSAWfP as described below:

\begin{itemize}
	\item The sequential flow between two tasks $A$ and $B$ can be expressed either by a production $p$ of the form $p:A \rightarrow B$, or by a production $q$ of the form $q:S \rightarrow A \fatsemi B$ in which $S$ is a (re)structuring symbol (see fig. \ref{routingWithLSPWfA}(a));
	\item The parallel flow between two tasks $A$ and $B$ is expressed using a production $p$ of the form $p:S \rightarrow A \parallel B$ (see fig. \ref{routingWithLSPWfA}(b));
	\item The alternative flow (choice) between two tasks $A1$ and $A2$ is expressed using two productions $p1$ and $p2$ such that $p1:S \rightarrow A1$ and $p2:S \rightarrow A2$; $S$ is a (re)structuring symbol expressing the fact that after "execution" of $S$, one must execute either task $A1$ or task $A2$ (see fig. \ref{routingWithLSPWfA}(c));
	\item Iterative routing (repetition) is expressed using recursive symbols. Thus the productions $p1:A \rightarrow B$, $p2:B \rightarrow C$ and $p3:B \rightarrow A$ express a potentially (transitive) iterative flow on the task $A$ (see fig. \ref{routingWithLSPWfA}(d)); $P_{7}:\; E1\rightarrow E1$ in the running example also expresses a direct iterative flow on $E1$ (see fig. \ref{fig:artefactsGlobaux}).
\end{itemize}

As defined in this paper, LSAWfP (like traditional workflow languages) is only interested in modelling the inherent characteristics of administrative processes: i.e. tasks and their scheduling, data produced and consumed by tasks, actors in charge of executing the tasks and their roles. Aspects related to the execution of processes are not taken into account. LSAWfP therefore differs from the majority of artifact-centric languages. For example, the proclet approach \cite{van2001proclets} requires the designer to take into account aspects related to communication (ports, channels, etc.) between the entities in charge of executing tasks (the proclets). The approach in \cite{lohmann2010artifact} imposes to express also artifacts' locations (artifacts can be mobile or not). The expressiveness of LSAWfP does not allow the designer to specify such aspects.

The language LSAWfP has several interesting features in addition to its expressiveness; in particular:
\begin{itemize}
	\item Its ability to represent scenarios using simple graphs (annotated trees) where existing languages use graphical formalisms (arbitrary graphs) that are more complex to implement;
	\item Its semi-declarative approach which would like the designer to describe the scheduling of a subset of tasks in a given production without expressing how the whole set of tasks is performed procedurally;
	\item Its usage of accreditations to model organizational aspects of process, making it particularly suitable for the modelling of administrative processes;
	\item Its modular approach using the scenario as the modelling unit;
	\item Its solid mathematical foundation mainly made up of a grammatical model that can be studied formally in the same way as Petri nets, while benefiting from the executable character that is recognized in such a tool.
\end{itemize}

In table \ref{tableau:comparison}, we make a preliminary comparative study of LSAWfP with some of the workflow languages (those that we find best related to LSAWfP) presented in section \ref{sec:relatedWork}. The criteria we have retained are as follows: 
\begin{itemize}
	\item The mathematical foundation: we specify the mathematical tool that serves as a basis for the language.
	\item The modelling paradigm: we clarify whether the language advocates a \textit{procedural} approach in which one says how the process is carried out or whether it advocates a \textit{declarative} approach in which, tasks are described simply and their sequences are specified using rules.
	\item The modelling formalism: does language propose a graphic formalism (\textit{graph-based}), a textual formalism using rules (\textit{rule-based}), or both ?
	\item The modelling unit: either \textit{process} to refer to the fact that the whole process is modelled at once, or \textit{user perception} to refer to the fact that the process is cut and modelled according to the perceptions of each actor on it, or \textit{scenario} to refer to the fact that each scenario of the process is modelled independently of the others.
	\item The highlighted conceptual model: we specify which of the informational, organizational and lifecycle models is emphasized by the considered language.
	\item The implementation: here we specify whether the language has an implementation or not.
	\item The independence regarding an execution technique: we specify if the language does not take into account aspects related to the technique and execution environment of the modelled processes.
\end{itemize}
\begin{table}[ht]
	\centering
	\caption{A comparison of LSAWfP with other workflow languages.}
	\label{tableau:comparison}
	{\scriptsize
	\begin{tabular}[t]{p{1.4cm}p{1.6cm}p{1.8cm}p{1.7cm}p{1.4cm}p{2cm}p{1.1cm}p{1.1cm}}
		\hline \hline
		\textbf{Language} & \textbf{Mathematical foundation} & \textbf{Modelling paradigm} & \textbf{Modelling formalism} & \textbf{Modelling unit} & \textbf{Highlighted conceptual model} & \textbf{Has an implementation ?} & \textbf{Is execution-context independent ?} \\
		\hline
		LSAWfP & Attributed Grammars & Semi-procedural (when designing scenario trees) and semi-declarative (when deriving rules for the GMWf) & Graph-based (annotated trees) and rule-based (the GMWf) & Scenario & All the models & No & Yes \\
		\hline
		BPMN \cite{BPMN} & Arbitrary Directed Graphs & Procedural & Graph-based & Process & Organizational and lifecycle models, but studies extending them to other processes perspectives do exist & Yes & Yes \\
		\hline
		YAWL \cite{van2013business} & Petri Nets & Procedural & Graph-based & Process & Lifecycle models, but studies extending them to other processes perspectives do exist & Yes & Yes \\
		\hline
		AWGAG \cite{badouel2015active} & Guarded Attributed Grammars & Declarative & Rule-based & User perception (modelling of active workspaces) & Informational and lifecycle models & No & No \\
		\hline
		Proclets \cite{van2001proclets} & Petri Nets & Procedural & Graph-based & User perception (modelling of proclets' classes) & Informational and lifecycle models & Not sure & No \\
		\hline \hline
	\end{tabular}}
\end{table}
These comparison criteria are not exhaustive and it is necessary to conduct a further study in order to better compare these languages. It would also be more appropriate to consider several other languages in this comparative study.

The LSAWfP language as presented here, is not perfect. The first criticisms we can make are the following: 
\begin{itemize}
	\item The organizational and informational models of LSAWfP formalized in this paper are quite simple. They are sufficient for the basic work carried out here to present the main concepts taken into account by this new language. However, it would be wise in further work, to better study organizations with administrative processes to improve these models.
	\item The expressiveness of LSAWfP has been analysed on basic routings and its applicability has been established on some examples of processes among which, the one described in this paper. Since LSAWfP also has a solid mathematical foundation, we have no doubt about its applicability in practice. However, it is essential to evaluate it in the modelling of larger processes in terms of number of actors, tasks, events, distribution (geographical distribution), etc. to definitively validate it.
\end{itemize}

\section{Ongoing Work on LSAWfP}
\label{sec:Discussion}
There is still a lot of work to be done to refine our models and achieve our goal of producing a complete workflow management infrastructure (a complete and solid workflow language, tools to assist in the design and validation of processes, a workflow execution environment, etc.). In this section, we present some of the work being currently done on LSAWfP.

~

\noindent\textbf{LSAWfP and workflow patterns}: one avenue we are currently exploring is that of measuring the expressiveness of LSAWfP in relation to workflow patterns \cite{van2012workflow}. This will allow us to characterize precisely the class of processes (beyond administrative processes) that this language can facilitate the modelling. To conduct this study, it is necessary to study in detail the different workflow patterns proposed in \cite{van2012workflow}, then find examples of processes highlighting these patterns and finally, find out how LSAWfP can help to model these processes.

~

\noindent\textbf{Towards a blockchain-like artifact-centric model of processes design and distributed execution based on cooperative edition of a mobile artifact}: 
we are also working to produce an artifact-centric model of business process management. In this model inspired by the work of Badouel et al. on cooperative editing \cite{badouelTchoupeCmcs, tchoupeAtemkeng2, tchoupezekeng2016reconciliation, tchoupeZekeng2017, zekengTchoupe2018, zekengndadji:hal-02375958}, the process tasks are executed by the various actors with the help of software agents that they pilot. These software agents are autonomous, reactive and communicate in peer to peer mode by exchanging an artifact (considered as "mobile") edited cooperatively. This mobile artifact is an annotated tree that represents the execution status of the process at each moment. For this purpose, it contains information on the tasks already executed, on the data produced during these executions and on the tasks ready to be executed.

\begin{figure}[ht!]
	\noindent
	\makebox[\textwidth]{\includegraphics[scale=0.28]{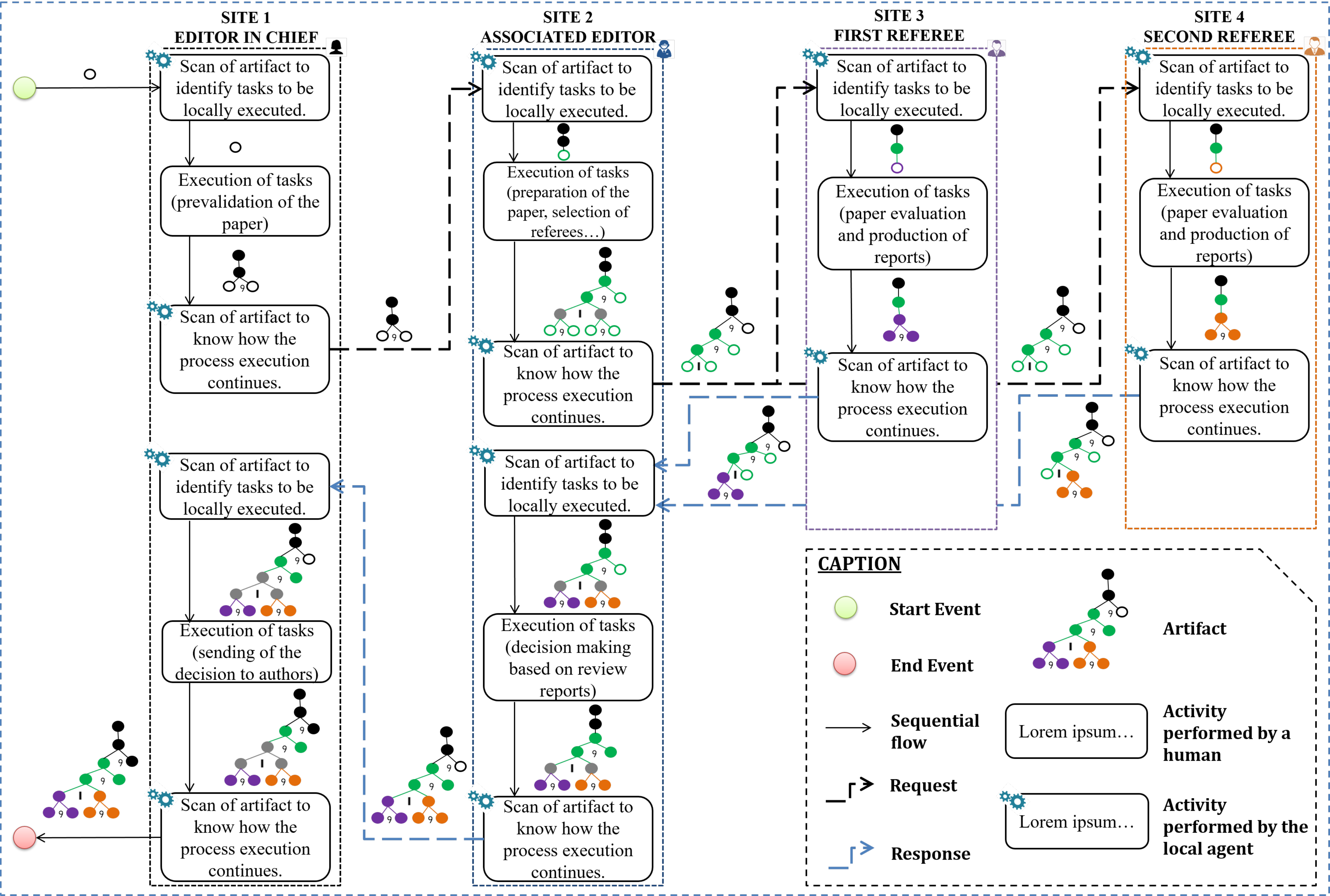}}
	\caption{An overview of the artifact-centric execution of the peer-review process.}
	\label{figOverviewExample}
\end{figure}

When the mobile artifact is received at a given execution site, the local agent executes an update protocol whose purpose is to reveal the tasks ready to be executed locally by the local actor. The execution of the tasks by the local actor is done using a specialized editor and can be assimilated to the edition of a structured document since its actions cause the received mobile artifact (the tree) to be updated, by expanding some of its leaf nodes into sub-trees and by assigning values to the "status" attributes of some other nodes. When all the tasks ready to be locally executed have been executed, the artifact is sent to other agents for further execution of the process if necessary.

To run the peer-review process described in section \ref{sec:runningExample} with the artifact-centric model being built, four agents controlled by four actors (the $EC$, the $AE$, the $R1$ and the $R$ agents) will be deployed. Figure \ref{figOverviewExample} sketches an overview of exchanges that can take place between those four agents. The scenario presented there corresponds to the nominal one in which the paper is pre-validated by the $EC$ and therefore, is analysed by a peer review committee. The artifact-centric execution is triggered on the $EC$'s site by introducing (in this site) an artifact reduced to its root node. During its transit through the system, this artifact grows. Note that there may be situations where multiple copies of the artifact are updated in parallel; this is notably the case when they are present on site 3 (first referee) and 4 (second referee).

\section{Conclusion}
\label{sec:Conclusion}
In this paper, we have proposed a new workflow language called LSAWfP which allows, through a simple grammar-based formalism, to specify administrative business processes. 
Like any traditional workflow language, LSAWfP allows to specify basic flows (sequential, parallel, alternative and iterative) that are generally found in workflow models;
particularly, it focuses on the modelling of each of the process scenarios using an artifact. Moreover, LSAWfP allows to model the main characteristics of business processes (their lifecycle, their informational and their organizational aspects); it also allows to address certain security aspects of administrative workflows. In fact, LSAWfP allows the workflow models designers, to simply express each actor's accreditations for each task in a process, by the means of a formalism inspired by that used in UNIX-like operating systems for the expression of users' rights. 
We also presented some of the work associated with LSAWfP that are currently in progress.
    
It would certainly be easier to handle LSAWfP if we had a (graphical) tool to assist in the design and validation of its instances. Such a tool could be built as a classical software engineering workbench through which the user would specify his processes by drawing trees and modifying the properties of their nodes, and then export valid specifications into dedicated formats after verification by simulation using a tool integrated into the workbench. Moreover, it seems equally important to more precisely describe the model for executing business processes specified in LSAWfP and briefly presented in section \ref{sec:Discussion}. In our opinion, this is just a few of the many studies that must be carried out following the one presented in this paper.

\begin{blindeddeclaration}{Authors' Information}
	\begin{itemize}
		\item \textbf{Milliam Maxime \textsc{Zekeng Ndadji}} is a PhD student in Computer Science at the University of Dschang (Cameroon), holding a Master of Science and a Bachelor of Science in Mathematics and Computer Science at the same university. His work focuses on the design of systems to support collaboration using formal tools such as grammars and automata.
		\item \textbf{Maurice \textsc{Tchoup\'e Tchendji}} holds a PhD in Software Engineering obtained in co-supervision at the Universities of Yaoundé I (Cameroon) and Rennes I (France); he is currently a senior lecturer and researcher at the University of Dschang (Cameroon). His work focuses on collaborative systems, XML databases, distributed systems, ad-hoc networks, improving the user experience through software localisation and machine learning.
		\item \textbf{Cl\'ementin \textsc{Tayou Djamegni}} obtained the Third Cycle Doctorate and the State Doctorate at the University of Yaounde I (Cameroon) in 1997 and 2005 respectively. He has been a senior lecturer and researcher at the University of Dschang (Cameroon) since 1996. His main interests concern algorithm parallelization, regular networks, concept formal analysis, distributed algorithms, distributed systems and the boolean satisfiability problem.
		\item \textbf{Didier \textsc{Parigot}} holds a PhD from the University of Paris-Sud (now Paris-Saclay, France) and is currently senior researcher on programming language at INRIA (Sophia Antipolis, France). He is interested in formal languages, distributed systems, service-oriented architectures, peer-to-peer computing, component-based software engineering, domain-specific languages and generative programming.
	\end{itemize}
\end{blindeddeclaration}

\begin{blindeddeclaration}{Authors' Contributions}
	\begin{itemize}
		\item \textbf{Milliam Maxime \textsc{Zekeng Ndadji}} suggested the idea of producing a workflow language based on an idea to produce a decentralised model of structured cooperative editing put forward by Tchoup\'e Tchendji Maurice and Didier Parigot. He then participated in the conception and formalisation of the said language, in the study of the latter with the help of several examples, in the writing and the proofreading of this paper. 
		\item \textbf{Maurice \textsc{Tchoup\'e Tchendji}} is at the root of the work from which the one in this paper is derived. He participated in the formalisation of the language presented here, in its illustration, in the writing and the proofreading of this paper.
		\item \textbf{Cl\'ementin \textsc{Tayou Djamegni}} is the co-supervisor of this work; he validated the various proposed mathematical tools, validated the examples that have been developed and contributed to the proofreading of this paper.
		\item \textbf{Didier \textsc{Parigot}} is (with Maurice \textsc{Tchoup\'e Tchendji}) at the base of the work that gave rise to this one; he is also co-supervisor of the work presented here. He has therefore validated the proposed mathematical tools and validated the examples that have been developed.
	\end{itemize}
\end{blindeddeclaration}

\begin{declaration}{Competing Interests}
	The authors declare that they have no competing interests.
\end{declaration}

\begin{declaration}{Funding}
	No funding was received for this project.
\end{declaration}

\begin{blindeddeclaration}{Editor}
	\textbf{Hector Florez}, Ph.D. \textit{Universidad Distrital Francisco Jose de Caldas, Colombia}\orcid{0000-0002-5339-4459}
\end{blindeddeclaration}    

\begin{blindeddeclaration}{Reviewers}
	\begin{itemize}
		\item \textbf{Raphael Gomes}, Ph.D. \textit{Instituto Federal de Goiás, Brazil}\orcid{0000-0002-6036-7678}
		\item \textbf{Jens Gulden}, Ph.D. \textit{Universiteit Utrecht, Netherlands}\orcid{0000-0003-4824-8569}
		\item \textbf{Ben Roelens}, Ph.D. \textit{Open Universiteit, Netherlands}\orcid{0000-0002-2443-8678}
	\end{itemize}
\end{blindeddeclaration}    

\references{bibliography}
\end{document}